\documentclass[preprint]{nature}
\usepackage{graphicx}
\usepackage{subfigure}
\usepackage{dcolumn}
\usepackage{bm}
\usepackage{array}
\usepackage{braket}
\usepackage{diagbox}
\usepackage{amsmath}
\usepackage{appendix}
\usepackage{multirow}
\usepackage{bm,color,bbm}
\usepackage{float}
\newcolumntype{.}{D{.}{.}{-1}}
\usepackage{amsopn}
\usepackage[colorlinks, linkcolor=blue, anchorcolor=black, urlcolor=red, citecolor=blue]{hyperref}
\usepackage{epstopdf}
\usepackage{booktabs}
\usepackage{enumerate}
\usepackage{siunitx}
\usepackage{upgreek}

\begin{document}
	\title{{High-rate quantum key distribution exceeding 110 Mb/s}}
	\author
	{Wei Li$^{1,2,\ast}$, Likang Zhang$^{1,2,\ast}$, Hao Tan$^{1,2}$, Yichen Lu$^{1,2}$, Sheng-Kai Liao$^{1,2,3}$, Jia Huang$^{4}$, Hao Li$^{4}$, Zhen Wang$^{4}$, Hao-Kun Mao$^{5}$, Bingze Yan$^{5}$, Qiong Li$^{5}$, Yang Liu$^{6}$, Qiang Zhang$^{1,2,3,6}$, Cheng-Zhi Peng$^{1,2,3}$, Lixin You$^{4}$, Feihu Xu$^{1,2,3,\star}$, Jian-Wei Pan$^{1,2,3,\star}$}
	
	\maketitle

	\begin{affiliations}
		\item Hefei National Research Center for Physical Sciences at the Microscale and School of Physical Sciences, University of Science and Technology of China, Hefei 230026, China
		\item Shanghai Research Center for Quantum Science and CAS Center for Excellence in Quantum Information and Quantum Physics, University of Science and Technology of China, Shanghai 201315, China
		\item Hefei National Laboratory, University of Science and Technology of China, Hefei 230088, China
		\item State Key Laboratory of Functional Materials for Informatics, Shanghai Institute of Microsystem and Information Technology, Chinese Academy of Sciences, Shanghai 200050, China
		\item School of  Cyberspace Science, Faculty of Computing, Harbin Institute of Technology, Harbin 150080, China
		\item Jinan Institute of Quantum Technology, Jinan, Shandong 250101, China
		\\
		$^\ast$These authors contributed equally. 
		\\ $^\star$e-mails: feihuxu@ustc.edu.cn; pan@ustc.edu.cn
	\end{affiliations}

	\begin{abstract}
		Quantum key distribution (QKD) can provide fundamentally proven security for secure communication. Toward application, the secret key rate (SKR) is a key figure of merit for any QKD system. So far, the SKR has been limited to about a few megabit-per-second (Mb/s). Here we report a QKD system that is able to generate key at a record high SKR of 115.8 Mb/s over 10-km standard fibre, and to distribute key over up to 328 km of ultra-low-loss fibre. This attributes to a multi-pixel superconducting nanowire single-photon detector with ultrahigh counting rate, an integrated transmitter that can stably encode polarization states with low error, a fast post-processing algorithm for generating key in real time and the high system clock-rate operation. The results demonstrate the feasibility of practical high-rate QKD with photonic techniques, thus opening its possibility for widespread applications.
	\end{abstract}

	\maketitle
	
	\paragraph{Introduction.}
	Quantum key distribution (QKD)\cite{BB84,ekert1991quantum} allows two remote parties to distill secret keys with information-theoretical security. During the past decades, QKD has drawn a lot of scientific attentions\cite{xu2019quantum,pirandola2020advances} as it not only provides quantum-secure cryptographic solutions but also insightful views to the bizarre quantum world. On the application front, increasing the secret key rate (SKR) of QKD is undoubtedly one of the pressing tasks as it not only allows more frequent key exchanges but also can provide services to a larger number of network users\cite{chen_integrated_2021} or high-data-rate applications such as critical infrastructure protection, sharing medical data and distributed storage encryption\cite{diamanti2016practical,sasaki2017quantum}.
	
	In the quest for high SKR, the system clock rate has been increased to gigahertz\cite{takesue2007quantum,lucamarini2013efficient,yuan2018,islam2017provably,boaron2018secure,grunenfelder2020performance}. Per-channel-use rate has been improved via low-error rate\cite{agnesi2020simple} and high detection efficiency\cite{islam2017provably,boaron2018secure}; advanced protocols such as twin-field QKD\cite{lucamarini2018overcoming} even hold the promise to beat the rate-loss law of repeaterless quantum communication. On the theory side, tight finite-key analysis\cite{tomamichel2012tight,lim2014concise,rusca2018finite} for acceptable accumulation times has been well studied. So far, 1 Mb/s SKR over 50-km fibre (10-dB loss) has been reported\cite{tanaka2012,lucamarini2013efficient,frohlich2017long}. Recently, a real-time SKR of 13.7 Mb/s over 2-dB loss emulated channel (equivalent to 10-km fibre) has been achieved\cite{yuan2018}. Nonetheless, these SKRs remain several orders of magnitude lower than current optical communication systems.
	
	Our primary interest is on the standard decoy-state QKD\cite{wang2005beating,lo2005decoy} as {it has been widely adopted in field implementations of metropolitan fibre links\cite{xu2019quantum,pirandola2020advances} and large-scale QKD networks\cite{chen_integrated_2021}}. Increasing the SKR faces several critical challenges including the transmitter, detection and post-processing. First, the high clock-rate implementation requires the stable and low-error modulation of the laser pulses and the encoding states, together with their driving electronics. Although gigahertz QKD systems are emerging in recent years, they often suffer from a large quantum bit error rate (QBER)\cite{lucamarini2013efficient,islam2017provably,yuan2018}. Second, high SKR needs the photon detection with both high efficiency and high count rate\cite{hadfield_single-photon_2009}. The superconducting nanowire single-photon detector (SNSPD) presents high efficiency and low noise\cite{marsili2013detecting,you2020}, but it has a long recovery time, thereby limiting the absolute SKR\cite{takesue2007quantum,islam2017provably,boaron2018secure,grunenfelder2020performance}; the fast gated InGaAs SPD can detect high photon flux, but its efficiency is limited for practical use\cite{korzh2015provably,comandar_quantum_2016}. Finally, the post-processing speed is a limiting factor for real-time key generation\cite{diamanti2016practical}. The sifted keys should be reconciliated in a fast and efficient way, and the privacy amplification has to support large input block size and high compression ratio.
	
	Here we address above challenges and report a polarization-encoding QKD system that is capable of generating a SKR of 115.8 Mb/s over 10-km standard fibre under the composable security against general attacks. For the source, we adopt an integrated modulator to realize the fast and stable modulations, the performance of which is optimized to produce an ultra-low QBER of 0.35\%. For the detection, we introduce the implementation of multi-pixel SNSPDs\cite{zhang201916} for both high-efficiency and high-rate photon detection. Our 8-pixel SNSPD has a maximum efficiency of 78\% at 1550-nm wavelength, which can detect 552 million photons per second at an efficiency of 62\%. For the classical post-processing, we adopt an enhanced Cascade reconciliation algorithm\cite{mao2022high} and a hybrid hash-based privacy amplification algorithm\cite{yan2022efficient} which achieves an average throughput of 344.3 Mb/s. Moreover, we develop high-speed electronics to operate the QKD system at 2.5-GHz clock rate, use an efficient polarization feedback control scheme, and adopt the protocol with 4-encoding states and 1-decoy state at the optimal SKR under the finite-key security\cite{lim2014concise,rusca2018finite}. All together, our QKD system enables a SKR enhancement about one order of magnitude over the previous BB84 record\cite{yuan2018}. The system robustness and stability are verified by a 50-hour continuous operation. We also show the feasibility to generate secret keys at long distances up to 328-km of ultra-low-loss fibre using polarization states. This verifies a 60-dB dynamic range of photon counting rate between short and long distance, thus proving the practicality of our QKD system for general application scenarios.
	
	\paragraph{Protocol.}
	In the BB84 protocol, the information is encoded in two conjugate basis, where the rectilinear basis (denoted as $Z$ basis) and the diagonal basis (denoted as $X$ basis) are used in the polarization dimension. {To increase the sifting efficiency, Alice and Bob preferentially choose an efficient version of the BB84 protocol with biased basis choice\cite{lim2014concise,rusca2018finite}.} That is, the $Z$ basis ($P_Z > 0.5$) is used to distill keys, while the bit information in $X$ basis is publicly announced to evaluate the information leakage to the eavesdropper. {This means that the count rates of $Z$ and $X$ bases are also asymmetric.} After sifting of the keys, Alice and Bob reconcile their keys to the same bits stream and distill the final secure keys through privacy amplification. The decoy state method is a standard approach to protect against photon-number-splitting attack\cite{wang2005beating,lo2005decoy}. Among the variants of decoy state method\cite{lim2014concise}, we adopt the 1-decoy state protocol\cite{rusca2018finite} due to the following two reasons. First, a vacuum state is not needed, thus putting less stringent requirement on the intensity modulator. Second, within a realistic finite-key size ($n_Z\le 10^8$) and when the QBER is low, the 1-decoy protocol gives higher SKRs at almost all distances (See Supplementary Section 5).
	
	Following ref.\cite{rusca2018finite}, the finite-key SKR (bits per second) under the composable security against general attacks can be bounded by:
	
	\begin{equation}\begin{aligned}
	\label{eq:skr}
	K=&\left[s_{Z, 0}^{l}+s_{Z, 1}^{l}\left(1-h\left(\phi_{Z}^{u}\right)\right)-f \cdot n_Z \cdot h\left(E_Z\right)\right.\\
	&\left.-6 \log _{2}\left(19 / \epsilon_{\sec }\right)-\log _{2}\left(2 / \epsilon_{\text {cor }}\right)\right] / t,
	\end{aligned}\end{equation}
	where $t$ is the data accumulation time, $ s_{Z, 1}^{l} \ (s_{Z, 0}^{l})$ is a lower bound on the number of single-photon contributions (vacuum-state contributions) in the sifted keys; $ \phi_{Z}^{u} $ is an upper bound on the single-photon phase-error rate; $h(\cdot)$ is the binary entropy function; $f$ is the efficiency of the error correction code; $ n_Z $ is the sifted key length in the $ Z $ basis; $ E_Z $ is the bit error rate in the $ Z $ basis; and $\epsilon_{\mathrm{sec}}$, $\epsilon_{\mathrm{cor}}$ are the secrecy and correctness parameters respectively.
	
\begin{figure}[!htb]
		\centering
		\includegraphics[width=0.8\linewidth]{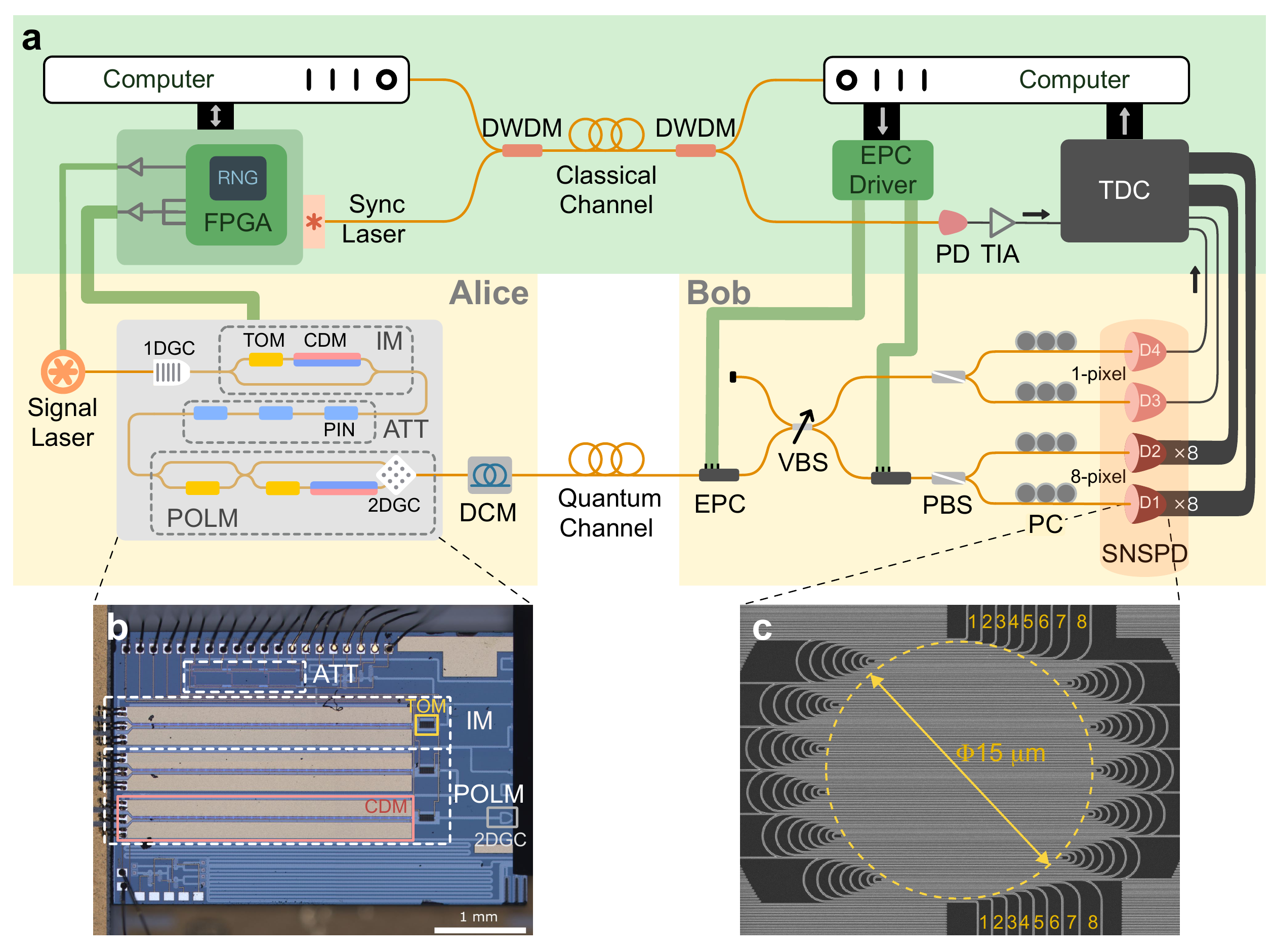}
		\caption{\textbf{a,} Depiction of the experimental setup. {The signal pulse is encoded in polarization by an integrated modulator and decoded by two 8-pixel SNSPDs (D1,D2) and two 1-pixel SNSPDs (D3,D4). The efficiencies of the detectors are balanced actively by the PC. The synchronization light is frequency-multiplexed with the classical communication channel. The post-processing unit is based on CPU platforms and the FPGA board used to drive electro-optic devices embeds a pseudo random binary sequence.} IM, intensity modulator; POLM, polarization modulator; ATT, variable attenuator; TOM, thermo-optic modulator; CDM, carrier-depletion modulator; 1DGC (2DGC), one-dimensional (two-dimensional) grating coupler; DCM, dispersion compensating module; VBS, variable beam splitter; EPC, electronic polarization controller; PBS, polarization beam splitter; SNSPD, superconducting nanowire single-photon detector; PC, polarization controller; RNG, random number generator; DWDM, dense wavelength-division multiplexer; PD, photodiode; TIA, transimpedance amplifier; TDC, time-to-digital converter. \textbf{b,} Microscopic view of the intergraded modulator chip. \textbf{c,} Scanning electron microscopy image of the 8-pixel SNSPD with interleaved nanowires. }
		\label{fig:setup}
	\end{figure}

	\paragraph{Setup.}
	To implement the protocol, we build a system as depicted in Fig.~\ref{fig:setup}a. A distributed-feedback laser (signal laser, {modeled Gooch \& Housego AA0701}) is gain-switched by a 120-ps pulse with carefully tuned pump intensity which enables the generation of 2.5-GHz phase-randomized pulse stream at 1550.12 nm. The waveforms of the driving signal and output light pulse are shown in Supplementary Fig. 7. Due to the amplified spontaneous emission process, each new pulse has a random phase, thus satisfying the phase randomization assumption for decoy state protocol\cite{yuan2014robust}. The light is coupled into a silicon photonic chip modulator through a one-dimensional grating coupler.
	
	The chip modulator (Fig.~\ref{fig:setup}b) modulates the intensity of decoy states via the intensity modulator, encodes the polarization states via the polarization modulator and attenuates the light to the single-photon level via the attenuator. The intensity modulator is realized by a Mach-Zehnder interferometer incorporating two types of phase modulators, i.e., thermo-optic modulator (TOM) and carrier-depletion modulator (CDM). The TOM is used for static phase bias while CDM is modulated dynamically. Due to the compact size and the precise temperature control, the intensity stability is within 0.1 dB even without the bias feedback throughout the experiments. The polarization modulator is structured by a Mach-Zehnder interferometer followed by a two-dimensional grating coupler. The use of CDM for high-visibility polarization modulation is challenging\cite{ma2016silicon,sibson2017integrated,wei2020high,avesani2021full}, because the variation of carrier induces the change in the refractive index thus causing loss dependence and the depletion of the carrier can reach a saturation point. We counter the effect of phase-dependent loss by optimizing the bias of the TOM in the first stage of polarization modulator, and optimize the design of CDM to increase its modulation efficiency. By precisely controlling the parameters and developing a homemade field-programmable gate array for electronic control, we are able to modulate four BB84 states dynamically at an average polarization extinction ratio of 23.7 dB (see Supplementary Section 1). This corresponds to an intrinsic QBER of 0.4\%.
	
	To avoid the pulse width from broadening due to the frequency chirp produced by the gain-switched laser, a dispersion compensating module is inserted before the quantum channel, loss of which is included in the attenuation of Alice. The link between Alice and Bob is constituted by standard telecom fibre spools (G.652). The synchronization signal at ITU channel 44 (1542.12 nm) is pulsed at 3 ns with a repetition rate of 152.6 kHz and is multiplexed with the classical communication using a 100 GHz dense wavelength division multiplexer.
	
	The detection setup passively selects the measurement basis with probability $ q_Z $ which is tuned to the same value as the basis sending probability $ p_Z $ by a variable beam splitter. The electronic polarization controller (EPC) in front of the variable beam splitter is used to align the polarization bases between Alice and Bob, while the second EPC is used to rotate from $ Z $ to $ X $ basis. The polarization feedback control is crucial for a stable polarization-encoding QKD systems\cite{xavier2008full}. We adopt the stochastic parallel gradient descent algorithm\cite{vorontsov1997adaptive} for the polarization feedback control where the QBERs in $ Z $ and $ X $ bases are used to feedback driving voltages of the EPC (see Methods).

\begin{figure}[!htb]
		\centering
		\includegraphics[width=0.9\linewidth]{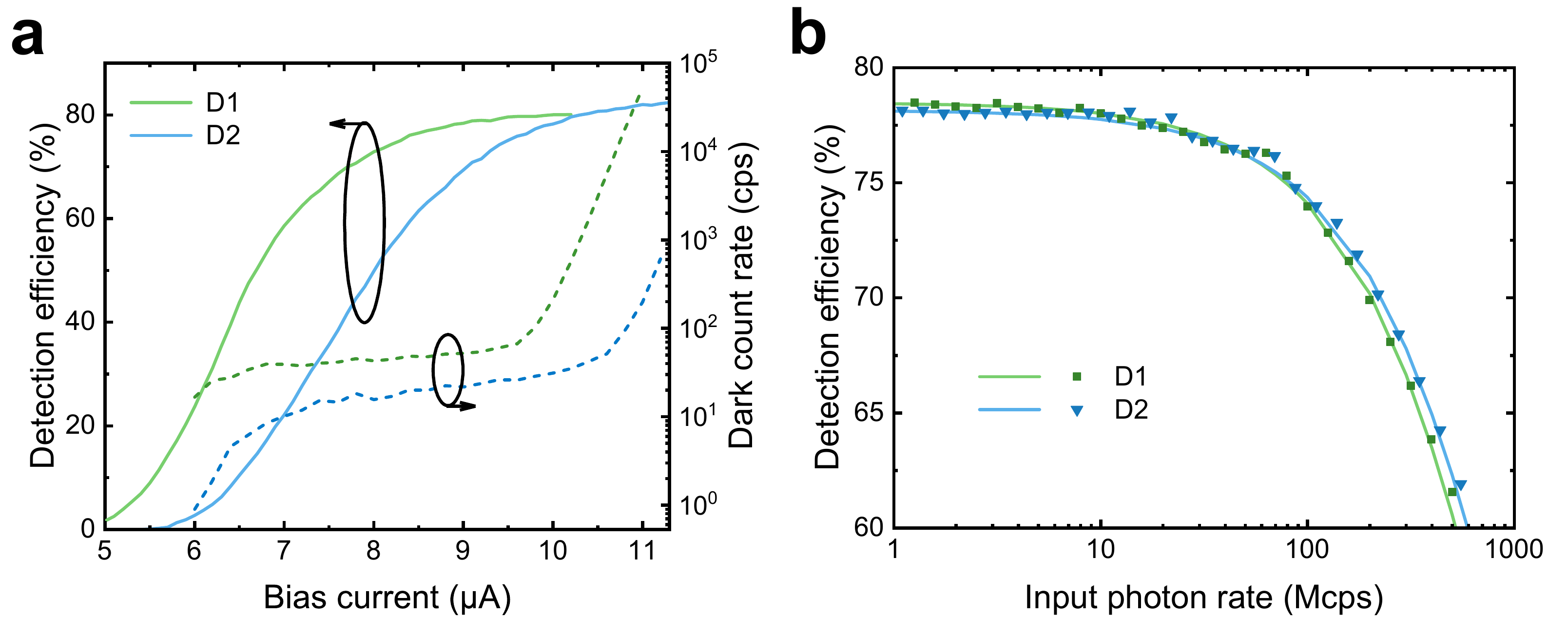}
		\caption{\textbf{a,} The total detection efficiency and dark count rate as functions of the bias current. \textbf{b,} Detection efficiency of two 8-pixel SNSPDs under different input photon flux.  The dots denote the measured results and the solid curves are the fitting result using the dead time model.}
		\label{fig:snspd}
	\end{figure}

	We implement multi-pixel SNSPDs for both high-efficiency and high-rate photon detection\cite{dauler2009photon,zhang201916}. Four NbN SNSPDs are enclosed in a cryogenic chamber and are cooled to 2.2 K. Benefiting from the asymmetric basis configuration, two 8-pixel SNSPDs are used for $ Z $ basis to accommodate high photon rate, while two 1-pixel SNSPDs are used for $X$ basis for standard photon detection. The 8-pixel SNSPD has 8 interleaved nanowires covering a circular active area 15 $\upmu$m in diameter, {and the nanowires are 75-nm wide (linewidth) with a lateral period (pitch) of 180 nm (Fig.~\ref{fig:setup}c). The linewidth and pitch are selected to ensure a near-unity absorption and considerable fabrication margin.} To increase the yield and uniformity of the interleaved nanowires, extended parallel nanowire structure outside the active area is designed to reduce the proximity effect in the process of electron-beam lithography. The electrical signal of each pixel is amplified and read out independently. As shown in Fig.~\ref{fig:snspd}a, the total efficiency is 78\% (78\%) and the total dark count is 52 (31) count/s when the bias current is set at 9 (10) $ \upmu $A for detector D1 (D2). At this bias current, the full width at half maximum of the timing jitter is about 60 ps for a single pixel. Fig.~\ref{fig:snspd}b shows the performance at high photon flux. To characterize the equivalent dead time of the multi-pixel SNSPD, we fit the count rate dependence on the input photon flux. The fitted dead time is 0.7 ns which is used in the key rate simulation and optimization. The sub-ns dead time assures the 8-pixel SNSPD with the maximum count rate of 342 Mcount/s. In contrast, for a dead time of 50 ns as a typical value for a single-pixel SNSPD\cite{hadfield_single-photon_2009}, it would saturate when the count rate exceeds 20 Mcount/s.
	
	The photon counting events are registered by a time-to-digital unit {(Time Tagger Ultra from Swabian Instruments)} which has a full width at half maximum timing jitter of 22 ps and a dead time of 2.1 ns channel-wise. We use the burst mode which can register 512 million events continuously at a rate of 475 Mcount/s. To distill the final secure key, the post-processing is performed on two Intel Core i7-10700 platforms communicating with each other via Gigabit Ethernet. The classical communication channel is consisted by 50-km fibre. Three steps are included in the post-processing: sifting, error reconciliation and privacy amplification. {Particularly, to realize high-speed post-processing, we design a high-performance solution of Cascade reconciliation for error correction\cite{mao2022high} involving two-way communications. Note however that a rigorous finite-key security analysis for the scenario of two-way error correction needs further study\cite{scarani2008quantum}. For high speed implementation of privacy amplification, we design a hybrid hash named multilinear-modular-hashing and modular arithmetic hashing\cite{yan2022efficient} (see Methods).}

\begin{figure}[!htb]
		\centering
		\includegraphics[width=0.8\linewidth]{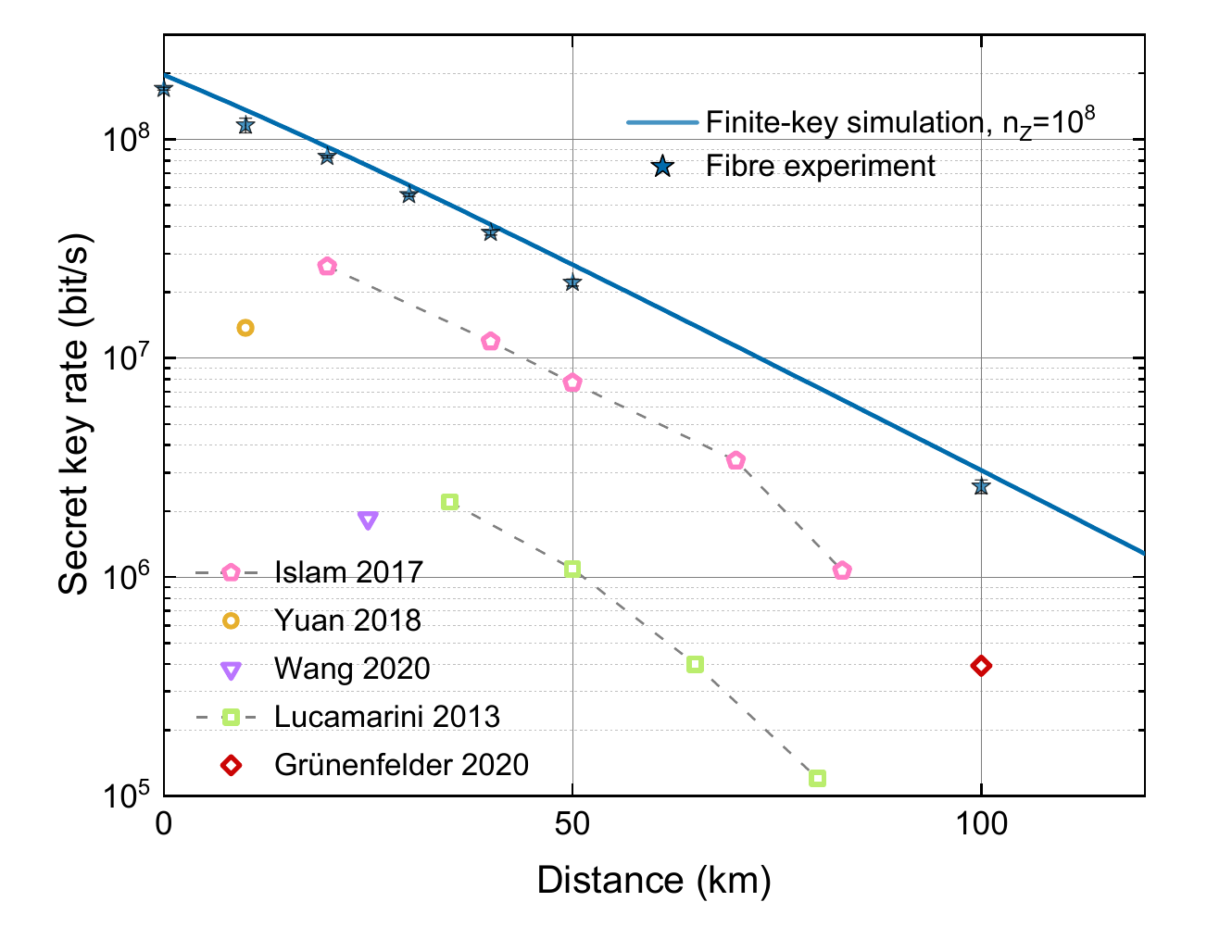}
		\caption{SKRs at different fibre distances. The solid line is the simulated SKRs using experiment parameters. The solid stars denote the experimental results using standard fibre spools. The SKRs are 170.3$ \pm $2.6 (sample size $ n $=7), 115.8$ \pm $8.9 ($ n $=3444), 83.2$ \pm $1.1 ($ n $=12), 55.7$ \pm $0.9 ($ n $=12), 37.4$ \pm $0.9 ($ n $=11), 22.2$ \pm $0.8 ($ n $=8), 2.6$ \pm $0.2 ($ n $=11) Mb/s respectively. The error bar denotes the standard deviation. The open symbols are the key-rate results of other high-rate QKD experiments implementing the high-dimensional (magenta pentagon)\cite{islam2017provably}, BB84 (yellow circle\cite{yuan2018}, green square\cite{lucamarini2013efficient} and red diamond\cite{grunenfelder2020performance}) and continuous-variable (purple inverted triangle\cite{wang2020high}) protocol respectively.}
		\label{fig:skr}
	\end{figure}

	\paragraph{Results.}
	Using the described setup, we perform a series of laboratory experiments from short to long distance transmissions using both standard fibre and ultra-low-loss fibre spools. We adopt secrecy and correctness parameters as $\epsilon_{\mathrm{sec}} = 10^{-10}$, $\epsilon_{\mathrm{cor}} = 10^{-15}$. The simulation and experimental results for the finite block size $n_Z=10^8$ are plotted in Fig.~\ref{fig:skr} (see Methods). The measured SKRs for 10-, 50- and 101-km standard fibre (loss of 2.2, 9.5, 19.6 dB) are 115.8$ \pm $8.9, 22.2$ \pm $0.8 and 2.6$ \pm $0.2 Mb/s with QBERs of 0.61$ \pm $0.10\%, 0.35$ \pm $0.05\% and 0.56$ \pm $0.11\%. See Supplementary Section 7 for detailed results. To highlight the progress entailed by our results, we compare our SKR along with recent high-rate QKD experiments in Fig.~\ref{fig:skr} and Tab.~\ref{tab:comparison}. Even considering high dimensional\cite{islam2017provably,lee2019large} and continuous variable\cite{wang2020high} QKD (using the local-local-oscillator protocol\cite{qi2015generating}), our work represents the highest SKR among the reported QKD systems.
	
	The increased QBER at short distance of 10 km is mainly caused by the pile-up of SNSPD pulses at high photon flux. We characterize the skew induced by the pile-up for each channel and apply a timing correction to the detection events based on the skew (see Supplementary Section 3). The correction is first-order, meaning that only adjacent pulses intervals are considered. After the correction, the QBER in $Z$ basis drops significantly from 7.01\% to 0.83\% for back to back scenario. The modulation error of the transmitter contributes 0.4\% to the QBER while the other is mainly contributed by the false registering of the detector at high photon rate.
	
	At long distance, we demonstrate 233$ \pm $112 bit/s SKR over 328-km ultra-low-loss fibre (55.1 dB channel loss) for a 29.5-hour run. The QBER increases to 2.8$ \pm $0.4\% which is mainly contributed by dark count noise (1.4\%) and polarization misalignment. We credit our successful polarization distribution over such long fibre to an advanced polarization compensation technique which uses strong pulses as feedback signals for the control algorithm. This result also represents the longest distance of fibre channel in polarization-encoding QKD systems\cite{xu2019quantum}. The security distance might be further extended by employing the filtering techniques to reduce the dark noise of SNSPD\cite{boaron2018secure}.
	
\begin{figure}[htbp]
		\centering
		\includegraphics[width=0.9\linewidth]{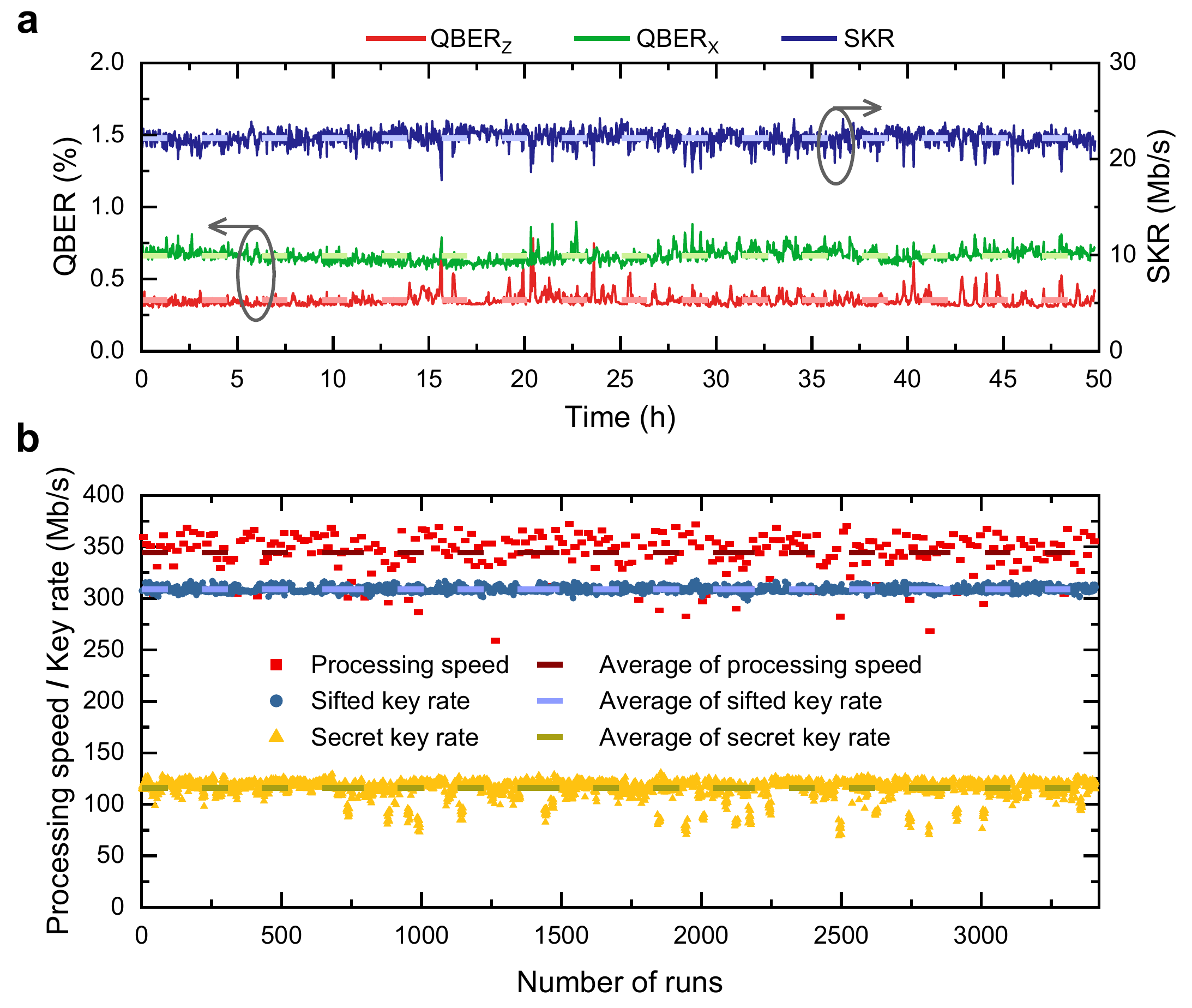}
		\caption{\textbf{a,} The QBER and SKR of 50-km standard fibre experiment measured for 50 hours continuous operations. The SKR is 22.2 Mb/s and the QBER in $ Z $ basis is 0.35\% on average. \textbf{b,} Key rates and post-processing speed for the runs of 10-km fibre. There are 3444 datasets of sifted keys in total, {and the size and the extracted key length of each dataset are $ 10^8 $ and 37,516,126 respectively.} Every 12 datasets are processed in parallel where the datasets share the same processing speed.}
		\label{fig:ppspeed}
	\end{figure}

	Fig.~\ref{fig:ppspeed}a shows the stability test result over 50-km fibre for 50 hours. This confirms the system robustness for continuous operations. {The small QBER spikes in the figure are mainly caused by the room temperature variations (see Supplementary Fig. 5). The variation caused disturbance to the polarization states transmitted in fibre spool which was not fully compensated. To validate the post-processing speed,} Fig.~\ref{fig:ppspeed}b shows the sifted and secret key rates for 3444 post-processed data blocks during 5-hour run under 10-km fibre channel. The rates are calculated Each data point was obtained when $ n_Z $ was accumulated to the size of $10^8$ bits. The processing speed of error correction and privacy amplification are shown on the same plot. An average processing speed of 344 Mb/s is achieved with an average error correction efficiency $ f $ of 1.053. Besides, the frame error rate is 0.021\% which has been considered into the calculation of the efficiency $ f $. {Importantly, the post-processing speed of error correction and privacy amplification has surpassed the average sifted key rate of 308.8 Mb/s for high-speed secret key extraction.}
	
	\paragraph{Discussion.}
	In summary, we have reported a QKD system capable of delivering secret keys at rates exceeding 115 Mb/s. To do so, we have developed a high-speed and stable QKD system, an integrated transmitter for low-error modulation, multi-pixel SNSPDs for high-rate detection and fast post-processing algorithms. Further SKR increase is possible using wavelength or spatial multiplexing technologies\cite{canas2017high,wengerowsky2018entanglement,bacco2019boosting}. {We note the recent important progress on high-rate CV-QKD\cite{wang_sub-gbps_2022,roumestan_high-rate_2021}, but the practical issues including the finite-key security proof against general attacks and the fast implementation of information reconciliation for discrete modulation CV-QKD remain to be resolved.} {Our implementation and security analysis do not consider the device imperfections. In practice, however, our system needs special care against the side-channel attacks\cite{xu2019quantum}. For high-speed QKD, polarization-dependent loss and intensity correlations are other important features to be characterized (Supplementary Section 6).}
	
	To our knowledge, our experiment is the first to show the superior performance of multi-pixel SNSPDs with interleaved nanowires for high-speed QKD. Although the multi-pixel SNSPD requires cryogenic cooling, our setup can be readily adopted in backbone QKD links\cite{chen_integrated_2021} so as to enhance the bandwidth and support more users. It is also suitable for an upstream quantum access network\cite{frohlich_quantum_2013} where a large number of transmitters multiplex a single detector. Besides, the silicon integrated modulator used in our setup can benefit users in cost, size and stability\cite{wang2020integrated}. Overall, the substantial increase of key rate demonstrated here could potentially open new opportunities in areas where data security is utmost important and bring QKD closer to widespread applications.
	
	\paragraph{Acknowledgments}
	The authors would like to thank Bing Bai, Ye Hong, Wei-Jun Zhang, Jun Zhang and Xiao Jiang for helpful discussions and assistance. This work was supported by National Key Research and Development Plan of China (Grant No. 2020YFA0309700), National Natural Science Foundation of China (Grant No. 62031024, 62071151), Innovation Program for Quantum Science and Technology (2021ZD0300300), Anhui Initiative in Quantum Information Technologies, Shanghai Municipal Science and Technology Major Project (Grant No. 2019SHZDZX01) and Chinese Academy of Sciences.  W.L. acknowledges support from the Natural Science Foundation of Shanghai (Grant No. 22ZR1468100). F. Xu acknowledge the support from the Tencent Foundation.
	
	\newpage
	\begin{table*}[htbp]
		\centering
		\caption{A list of high-rate QKD experiments. CR, clock rate; DE, detector efficiency; SKR, secret key rate; PP, post-processing; CV, continuous variable; BHD, balance homodyne detector. $^{\dagger}$Emulated attenuation, fibre channels otherwise.} \label{tab:comparison}
		\vspace{4mm}
		\begin{tabular}{@{}lllllllll@{}}
			\hline
			\textbf{Reference} &
			\textbf{Protocol} &	
			\begin{tabular}[c]{@{}l@{}}\textbf{CR}\\ \textbf{(GHz)}\end{tabular} &
			\begin{tabular}[c]{@{}l@{}}\textbf{QBER}\\ \textbf{(\%)}\end{tabular} &
			\begin{tabular}[c]{@{}l@{}}\textbf{DE}\\ \textbf{(\%)}\end{tabular} &
			\begin{tabular}[c]{@{}l@{}}\textbf{Detector}\\ \end{tabular} &	
			\begin{tabular}[c]{@{}l@{}}\textbf{Loss}\\ \textbf{(dB)}\end{tabular} &
			\begin{tabular}[c]{@{}l@{}}\textbf{SKR}\\ \textbf{(Mb/s)}\end{tabular} &
			\begin{tabular}[c]{@{}l@{}}\textbf{PP}\\ \end{tabular}			  \\ \midrule
			Lucamarini et al.\cite{lucamarini2013efficient} & Decoy BB84& 1 & 4.26 & 20 & InGaAs &7.0    & 2.20 & No\\
			Yuan et al.\cite{yuan2018}          & Decoy BB84 &1&  3.0      & 31 & InGaAs& 2.0$^{\dagger}$ & 13.72 & Yes\\
			Gr\"{u}nenfelder et al.\cite{grunenfelder2020performance} & Decoy BB84 & 5 & 1.9 & 80 & SNSPD & 20.2 & 0.39 & No \\
			Islam et al.\cite{islam2017provably}& High dimension & 2.5 & 4.0 & 70 &SNSPD& 4.0$^{\dagger}$  & 26.2 & No \\							
			Wang et al.\cite{wang2020high}      & Gaussian CV & 0.1 & N/A & 56 & BHD & 5.0 &  1.85 & No\\		
			This work                & Decoy BB84 & 2.5 & 0.61 & 78 &SNSPD & 2.2 & 115.8 & Yes\\ \hline
		\end{tabular}
	\end{table*}
	
	\clearpage
	
	\clearpage
	\paragraph{References}

	\newpage
	\begin{methods}
		\subsection{Integrated modulator.}
		Fig.~\ref{fig:setup}a,b shows all functional block on the integrated chip for intensity and polarization modulation. The chip fabricated by the commercial foundry process in a $4.8\times3~\mathrm{mm}^2$ size wafer is packaged with a thermoelectric cooler. The TOM is designed to have an ohmic resistance of 680 $ \Upomega$. While three CDMs are 3.2-mm-long resulting in an electronic bandwidth of 21 GHz and a $ V_{\pi} $ of 4.7 V. Detailed performance analysis related to QKD application can be found in Supplementary Section 1. Due to the saturation of the modulator, 8-V peak-to-peak voltage is needed to produce $ 3 \pi / 2 $ phase change. A homemade field-programmable gate array board is used to generate the driving pulses (See Supplementary Section 2). Four BB84 states $|\psi\rangle=\left(|H\rangle+e^{i \theta}|V\rangle\right) / \sqrt{2}, \theta \in\{0, \pi / 2, \pi, 3 \pi / 2\}$ are prepared, where $ \theta $ is the phase modulated by the CDM in front of the two-dimensional grating coupler and $ |H\rangle $ ($ |V\rangle $) corresponds to the polarization state in the upper (lower) arm of the 2DGC.
		
		The variable attenuator is composed by a p-i-n junction working on carrier injection. A 3-V voltage can induce 38-dB loss variation. The large power consumed by forward-bias diode needs to be taken care. For a 3-V bias, 230-mA diode current would result in 0.69-W power consumption. This would potentially overload the thermoelectric cooler and change the working condition of other on-chip modulators. Thus we use three cascaded diodes to reduce the voltage applied to each one and reduce the total power consumed.
		
		\subsection{Post processing.}
		For the error reconciliation, a Medium-Efficiency mode is used for block-length setting and 12 threads are run for parallel computing. In each thread, 100 processing units are applied, and each unit processes a frame of length $ L=64 $ kb. Once a processing unit completes the task of error correction, the verification of 64-bit cyclic redundancy check is operated. The corrected frame is transferred to the privacy amplification module. If the verification is negative, the frame will be revealed and the information leakage is accounted in the reconciliation efficiency. For the privacy amplification, we set the block number $ k=2 $, that can support a maximum compression ratio of $ 50\% $. The length of a single block $ \gamma $ is set to $ 57,885,161 $, and the corresponding input data size of privacy amplification is $ N=\gamma \times k =115,770,322 $, which is larger than the finite key size of $ 10^8 $.
		
		\subsection{Polarization compensation.}
		In our experiments, we adopt the stochastic parallel gradient descent algorithm for polarization feedback control (Supplementary Section 4). The algorithm uses the QBER in $ Z $ and $ X $ bases as error signals (objective function) to feedback driving voltages of the EPC in front of the variable beam-splitter possessed by Bob. The controller consists of three fibre squeezers controlled by direct-current voltages applied on piezoelectric elements. The squeezers are aligned \ang{0}, \ang{45} and \ang{0} respectively. Alice sends sufficient calibration signals in the $ Z $ and $ X $ bases using the same sending probability as quantum signals. Bob collects the QBER of the calibration sequence and uses it as the feedback signal. The polarization basis of Alice and Bob can be aligned by keeping the two QBER values low. During the experiment over 328-km-long fibre, we use strong calibration pulses which is 12.9 dB larger than signal pulses in intensity and the accumulation time is 0.5 s. The ratios of calibration pulses are 1/8 and 1/256 for 328-km experiments and the others respectively.
		
		\subsection{Finite-key simulation.}
		For dedicated fibre, the observed yield and error rate per photon pulse prepared in basis \(B\) and intensity \(\mu_{k}\) is given by:
		
		\begin{align}
		D_{B,k}&=1-\big(1-2p_{\rm dc}\big) e^{-\mu_{k}\eta_{\rm sys}q_{_B}};\\
		Q_{B,k}&=D_{B,k}\big(1+p_{\rm ap}\big) \\
		Q_{B,k}E_{B,k}&=p_{\rm dc}+e_{\rm mis}\big(1-2p_{\rm dc}\big)\big(1-e^{-\mu_{k}\eta_{\rm sys}q_{_B}}\big)+\frac{1}{2}p_{\rm ap}D_{B,k}
		\end{align}
		where \(B=Z,X\) stands for the basis, \(\mu_{k}\) is the \(k\)-th intensity in the protocol while \(k=1,2\) stands for the intensity index, \(q_{B}\) is Bob's probability of selecting basis \(B\) passively, \(\eta_{\rm sys}\) is the overall transmittance including the channel transmittance $\eta_{\rm ch}$ and the efficiency of Bob's detection system $\eta_{\rm Bob}$. Thereafter, using \(Q_{B,k}\) and \(Q_{B,k}E_{B,k}\) generated from the above model as input, the bounds of single-photon (vacuum-state) contributions \(s_{Z, 1}^{l}\) (\(s_{Z, 0}^{l}\)) as well as the single-photon phase error rate \(\phi_{Z}^{u}\) can be estimated by finite-key analysis and decoy-state methods. The final secret key rate is given by equation (\ref{eq:skr}) in the main text.
		
		In the simulation, we assume the probability is the same for Alice and Bob to choose Z basis: \(p_{Z}=q_{Z}\). The simulation is based on a fixed finite raw key length \(n_{Z}=10^8\). The overall efficiency on Bob’s side is $\eta_{\rm Bob}=56.08\%$, including the detector efficiency and internal losses of Bob’s apparatus. The detector efficiency is set to 65$\%$, the same as the lowest efficiency of the four detectors due to the security requirement, and the overall internal loss of Bob’s apparatus is measured to be 0.64 dB. The dark count probability of one detector is $p_{\rm dc} = 10^{-8}$ per pulse, the dead time of one detector is $t_{\rm dt} =0.7$ ns, and the misalignment error rate is $e_{\rm mis} = 0.4\%$. The channel transmittance from Alice to Bob is $\eta_{\rm ch} = 10^{-0.19L/10}$, in which L is the fibre length of the quantum channel. The security parameters are set to be $\epsilon_{\rm sec} = 10^{-10}$ and $\epsilon_{\rm cor} = 10^{-15}$. Based on the model and parameters, we can optimize the parameters $(p_{Z}, \mu_{1}, \mu_{2}, P_{\mu_{1}}, P_{\mu_{2}})$ by maximizing the SKR for any channel loss, where \(P_{\mu_{k}}\) is the probability of sending \(\mu_{k}\).
		
	\end{methods}		
\end{document}